\documentclass[preprint,superscriptaddress,nofootinbib,11pt]{revtex4-1}
\usepackage{graphicx, epsfig}
\usepackage{xspace}
\usepackage{amsmath,amssymb}
\usepackage{hyperref}
\usepackage[utf8]{inputenc}

\def\nn{\nonumber}

\newcommand{\zt}{$\mathbb{Z}_2$}

\newcommand{\veff}{V_{\text{eff}}}
\newcommand{\vtree}{V^{(0)}}
\newcommand{\vone}{V^{(1)}}
\newcommand{\vtwo}{V^{(2)}}
\newcommand{\lagr}{\mathcal{L}}
\newcommand{\vphys}{v_\text{phys}}
\newcommand{\delo}{\delta^{(1)}}
\newcommand{\delt}{\delta^{(2)}}

\newcommand{\hc}{\text{h.c.}}
\DeclareMathOperator{\llog}{\overline{\text{log}}}

\DeclareMathOperator{\dlamb}{\mathcal{D}_3}
\def\msbar{{\ensuremath{\overline{\rm MS}}}\xspace}

\begin{document}

\preprint{OU-HET-1001}

\title{
 On two-loop corrections to the Higgs trilinear coupling in models with extended scalar sectors
}

%%%%%%%%%%%%%%%%%%%%%%%%
\author{Johannes~Braathen}
\email{braathen@het.phys.sci.osaka-u.ac.jp}
\affiliation{
Department of Physics,
Osaka University,
Toyonaka, Osaka 560-0043, Japan
}
%%%%%%%%%%%%%%%%%%%%%%%%

%%%%%%%%%%%%%%%%%%%%%%%%
\author{Shinya~Kanemura}
\email{kanemu@het.phys.sci.osaka-u.ac.jp}
\affiliation{
Department of Physics,
Osaka University,
Toyonaka, Osaka 560-0043, Japan
}
%%%%%%%%%%%%%%%%%%%%%%%%

\begin{abstract}
 We investigate the possible size of two-loop radiative corrections to the Higgs trilinear coupling $\lambda_{hhh}$ in two types of models with extended Higgs sectors, namely in a Two-Higgs-Doublet Model (2HDM) and in the Inert Doublet Model (IDM). We calculate the leading contributions at two loops arising from the additional (heavy) scalars and the top quark of these theories in the effective-potential approximation. We include all necessary conversion shifts in order to obtain expressions both in the \msbar and on-shell renormalisation schemes, and in particular, we devise a consistent ``on-shell'' prescription for the soft-breaking mass of the 2HDM at the two-loop level. We illustrate our analytical results with numerical studies of simple aligned scenarios and show that the two-loop corrections to $\lambda_{hhh}$ remain smaller than their one-loop counterparts, with a typical size being $10-20\%$ of the one-loop corrections, at least while perturbative unitarity conditions are fulfilled. As a consequence, the existence of a large deviation of the Higgs trilinear coupling from the prediction in the Standard Model, which has been discussed in the literature at one loop, is not altered significantly.
\end{abstract}

\maketitle

%%%%%%%%%%%%%%%%%%%%%%
%%%  Introduction  %%%
%%%%%%%%%%%%%%%%%%%%%%

\section{Introduction}

Although the Standard Model (SM) particle spectrum has been completed by the discovery of a 125-GeV Higgs particle at the CERN LHC \cite{Chatrchyan:2012xdj,Aad:2012tfa}, no sign of any new Physics has been found so far, and direct searches of non-SM particles are currently putting increasingly stringent bounds on parameter spaces of Beyond-the-Standard-Model (BSM) theories. At the moment, the measured properties of the Higgs boson appear to be in close agreement with their SM predictions, which tends to indicate that new Physics is either heavy or made difficult to observe by some mechanism such as alignment \cite{Gunion:2002zf} -- which is defined as the situation where the Higgs vacuum expectation value (VEV) is colinear in field space with one (often the lightest) of the CP-even Higgs mass eigenstates. As a consequence, in aligned scenarios of BSM models, the coupling constants of the 125-GeV Higgs boson are equal at \textit{tree level} to those in the SM, and deviations can only arise via radiative corrections. However, in models with extended Higgs sectors, some of the couplings of the SM-like Higgs boson can deviate significantly from the SM case because of non-decoupling loop effects involving the additional scalar states of the theory, as was found first in Refs.  \cite{Kanemura:2002vm,Kanemura:2004mg}. Among these is the Higgs trilinear coupling $\lambda_{hhh}$, on which we will focus in this letter. 

This coupling is especially important because it participates in the determination of the shape of the Higgs potential, and in turn the type and strength of the 
electroweak phase transition (EWPT). In particular, it has been shown in Refs. \cite{Grojean:2004xa,Kanemura:2004ch} that large -- $\mathcal{O}(20-30\%)$ or more -- deviations in $\lambda_{hhh}$ from its SM prediction are required for the EWPT to be of strong first order, which is necessary for the scenario of electroweak baryogenesis (EWBG) \cite{Sakharov:1967dj,Kuzmin:1985mm,Cohen:1993nk} to be successful.

Current experimental limits on the Higgs trilinear coupling, obtained via searches for Higgs pair production, are at 95\% confidence level $-5.0<\lambda_{hhh}/\lambda_{hhh}^\text{SM}<12.1$ from ATLAS \cite{Ferrari:2018akh} (see also \cite{Aaboud:2018ftw}) and $-11<\lambda_{hhh}/\lambda_{hhh}^\text{SM}<17$ from CMS \cite{Sirunyan:2018iwt} (see also \cite{Sirunyan:2018two}). As for further measurement prospects, the HL-LHC with 3$\text{ ab}^{-1}$ of integrated luminosity could reach \cite{DiVita:2017vrr} $0.1<\lambda_{hhh}/\lambda_{hhh}^\text{SM}<2.3$, while at the HE-LHC (a possible 27-TeV upgrade of the LHC) it might be possible to achieve the limits $0.58<\lambda_{hhh}/\lambda_{hhh}^\text{SM}<1.45$ using 15$\text{ ab}^{-1}$ of data \cite{Homiller:2018dgu,Cepeda:2019klc}. In the case of lepton colliders, the ILC operating at 250 GeV cannot access directly the Higgs trilinear coupling \cite{Fujii:2017vwa}, but its extension to 500 GeV (1 TeV) could measure $\lambda_{hhh}$ to a precision of 27\% (10\%) \cite{Fujii:2015jha} using all available datasets; independently CLIC running at 1.4 and 3 TeV could obtain a result to $\sim20\%$ precision (at 68\% confidence level) \cite{DiVita:2017vrr}. Finally, at a possible 100-TeV hadron collider, one could attain a level of accuracy as good as $\sim 5-7\%$ when using 30$\text{ ab}^{-1}$ of data \cite{Goncalves:2018yva,Chang:2018uwu}. 

Radiative corrections to the Higgs trilinear coupling were first investigated at the one-loop order in the SM and the minimal supersymmetric SM (MSSM) in Refs.~\cite{Barger:1991ed,Hollik:2001px,Dobado:2002jz}. Moreover, one-loop effects have also been investigated for various (non-supersymmetric) BSM theories with extended Higgs sectors -- namely with additional singlets \cite{Kanemura:2015fra,Kanemura:2016lkz,He:2016sqr,Kanemura:2017wtm}, doublets \cite{Kanemura:2002vm,Kanemura:2004mg,Kanemura:2015mxa,Arhrib:2015hoa,Kanemura:2016sos,Kanemura:2017wtm}, or triplets \cite{Aoki:2012jj} -- and most of these results are now available in the program \texttt{H-COUP} \cite{Kanemura:2017gbi}.  
Since Refs.~\cite{Kanemura:2002vm,Kanemura:2004mg} it is known that, in the non-decoupling regime, the dominant one-loop BSM corrections to the Higgs trilinear coupling can cause a deviation of $\lambda_{hhh}$ by several tens of or even a hundred percent from its SM prediction, while still verifying the criterion of tree-level unitarity \cite{Lee:1977eg}. After encountering such large effects at one loop, one may at first ask whether perturbativity is still preserved. This is indeed the case because the one-loop expressions are not a perturbation of the tree-level formula, and instead involve new parameters that only enter the calculation at loop level. However, it remains natural to enquire about the situation at two loops: $i.e.$, whether new effects as large as $\mathcal{O}(100\%)$ may add up with the one-loop ones, or whether contributions at two loops stay smaller than at one loop. 

The first study of leading two-loop corrections in a model exhibiting non-decoupling effects was performed in Ref.~\cite{Senaha:2018xek} for the case of the Inert Doublet Model (IDM), and indicated that two-loop corrections enhance the Higgs trilinear coupling by a few percent and slightly weaken the first-order EWPT. We should also mention anterior works in the context of supersymmetric models, motivated by the need for a consistent theoretical determination of the Higgs mass(es) and trilinear coupling: these are namely Refs.~\cite{Brucherseifer:2013qva} and \cite{Muhlleitner:2015dua} where the leading $\mathcal{O}(\alpha_t\alpha_s)$ SUSY-QCD corrections to $\lambda_{hhh}$ were computed in the MSSM and NMSSM respectively, and their effects were found to be of the order of 10\%. 

In this work, we continue along this line of research and investigate the possible size of two-loop corrections to $\lambda_{hhh}$ both in the IDM and in an aligned scenario of a Two-Higgs-Doublet Model (2HDM), using the effective-potential method. For the former, we include new scalar diagrams that were overlooked in Ref. \cite{Senaha:2018xek}, while for the latter the expressions that we obtain constitute the first results in the literature for the 2HDM and in general for two-loop diagrams involving both heavy Higgs scalars and top quarks. Moreover, we find the need for a careful treatment of the renormalisation of the soft-breaking mass $M$ of the 2HDM and we therefore devise a new prescription ensuring explicitly the decoupling of our expressions in terms of on-shell-renormalised parameters.  
We will restrict our attention here to the two-loop BSM contributions to $\lambda_{hhh}$ from the additional states in the extended Higgs sectors.

%%%%%%%%%%%%%%%%%%%%%%%%%%%%%%%%%%%
%%%%%%%%%%%   Models   %%%%%%%%%%%%
%%%%%%%%%%%%%%%%%%%%%%%%%%%%%%%%%%%

\section{Models}
\label{SEC:Models}
We here briefly recall our conventions for the 2HDM and the IDM. For more complete reviews of these models, see Refs.~\cite{Gunion:1989we, Branco:2011iw} for the 2HDM and Refs.~\cite{Barbieri:2006dq,Goudelis:2013uca} for the IDM. 

\subsection{Two-Higgs-Doublet Model}
\label{SEC:models_2HDM}
The first type of model that we consider is a CP-conserving 2HDM, defined in terms of two $SU(2)_L$ doublets $\Phi_1$, $\Phi_2$ of hypercharge $1/2$. To avoid flavour changing neutral currents that are strongly constrained experimentally, we impose a \zt~symmetry under which the two doublets of the theory transform as $\Phi_1\to\Phi_1$ and $\Phi_2\to-\Phi_2$ \cite{Glashow:1976nt}, but which is softly broken by a mass term ($m_3^2$) in the potential. 
We follow the conventions of Ref.~\cite{Kanemura:2004mg} and write the tree-level scalar potential as
\begin{align}
 \vtree_\text{2HDM}=&\ m_{1}^2 |\Phi_1|^2 + m_{2}^2 |\Phi_2|^2 -  m_3^2 (\Phi_2^\dagger\Phi_1 + \Phi_1^\dagger\Phi_2)\\ 
        & +\frac{\lambda_1}{2}|\Phi_1|^4 +\frac{\lambda_2}{2}|\Phi_2|^4 + \lambda_3 |\Phi_1|^2|\Phi_2|^2 +\lambda_4 |\Phi_2^\dagger\Phi_1|^2  +\frac{\lambda_5}{2} \Big((\Phi_2^\dagger\Phi_1)^2 + \text{h.c.}\Big)\,.\nn
\end{align}
Because we assume that CP is conserved, all mass parameters $m_i^2$ and quartic coupling constants $\lambda_i$ are real. We choose then to expand the doublets $\Phi_1$ and $\Phi_2$ as $\Phi_i=(\phi_i^+,\ \phi_i^0/\sqrt{2})$. Depending on the parameters of the Lagrangian, the neutral components of the doublets may acquire non-zero (real) vacuum expectation values (VEVs), which are denoted $v_i\equiv \langle\phi_i^0\rangle$ and verify $v_1^2+v_2^2=v^2$, where $v\simeq 246$ GeV. 

Assuming that both $v_1$ and $v_2$ are non-zero, two dimensionful parameters -- typically $m_1^2$ and $m_2^2$ -- can be eliminated using the tadpole equations (see $e.g.$ eqs~(9)-(10) in \cite{Kanemura:2004mg} for their tree-level expressions). Seven free parameters then remain in the scalar sector: $m_3^2$, $\lambda_i$ ($i=1-5$) and the ratio of the two VEVs $v_2/v_1=\tan\beta$. The latter defines an angle $\beta$ that diagonalises the charged and CP-odd Higgs mass matrices at tree-level, while for the CP-even Higgs mass matrix a second mixing angle $\alpha$ needs to be introduced. We can then obtain the charged and neutral components of the $SU(2)_L$ doublets in terms of mass eigenstates as
\begin{equation}
 \begin{pmatrix}
  \phi_1^+\\
  \phi_2^+
 \end{pmatrix}
= R_\beta \begin{pmatrix}
  G^+\\
  H^+
 \end{pmatrix}\,,\qquad
 \begin{pmatrix}
  \phi_1^0\\
  \phi_2^0
 \end{pmatrix}
= v\begin{pmatrix}
     c_\beta\\
     s_\beta
   \end{pmatrix}+
   R_\alpha \begin{pmatrix}
     H\\
     h
   \end{pmatrix}+i R_\beta \begin{pmatrix}
  G\\
  A
 \end{pmatrix}\,,
\end{equation}
with $R_x\equiv \left(\begin{smallmatrix}
                      \cos x & -\sin x\\ \sin x & \cos x
                     \end{smallmatrix}
\right)$, and where $h$ and $H$ are CP-even Higgs bosons, $A$ is a CP-odd Higgs boson, and $H^+$ is a charged Higgs boson. In addition, $G$ and $G^+$ are respectively the neutral and charged Goldstone bosons associated with electroweak symmetry breaking (EWSB). 
 
It is common to replace the Lagrangian mass parameter $m_3^2$ by the soft-breaking mass $M^2\equiv 2m_3^2/s_{2\beta}$, and to express the five quartic couplings in terms of the four scalar mass eigenvalues and the mixing angle $\alpha$, using tree-level relations given for example in equations (26)-(30) of Ref.~\cite{Kanemura:2004mg}. In this respect, it is important to emphasize that the mass eigenvalues in these equations should be interpreted as the \textit{tree-level} ones -- otherwise radiative corrections would need to be taken into account to obtain the relation between quartic couplings and loop-level mass eigenvalues (see for example Refs.~\cite{Kanemura:2017wtm,Kanemura:2015mxa,Krause:2016oke,Basler:2016obg,Braathen:2017izn, Braathen:2017jvs,Basler:2017uxn}).

To ensure compatibility with experimental constraints, we will throughout this letter consider the so-called \textit{alignment limit}. This limit is defined by the requirement that one of the CP-even Higgs mass eigenstates is aligned in field space with the full Higgs VEV $v$ \cite{Gunion:2002zf}. Additionally, we want the SM-like state to be the lightest eigenstate $h$, which in terms of mixing angles implies $s_{\beta-\alpha}=1$, or equivalently $\alpha=\beta-\pi/2$. The tree-level couplings of $h$ to other particles are then equal to their SM values, and in particular $\lambda_{hhh}^{(0)}=3m_h^2/v^2$. 

Furthermore, in the alignment limit and for $m_h\ll m_\Phi$, we can obtain simple expressions for the field-dependent tree-level masses of the additional scalars 
$\Phi=H,A,H^\pm$ as
\begin{equation}
 m_\Phi^2(h)\simeq M^2+\frac{m_\Phi^2-M^2}{v^2}(v+h)^2 \,.
\end{equation}

Finally, it should be noted that we neglect throughout this letter contributions from quarks other than the top and from leptons, so there is no need to specify the type of fermion couplings in our setting. Indeed, at tree-level, the couplings of the top quark to the scalar sector are the same in all types of 2HDMs, and its field-dependent mass is $m_t^2(h)=y_t^2s_\beta^2(v+h)^2/2$. 

\subsection{The Inert Doublet Model}
\label{SEC:models_IDM}
The IDM \cite{Deshpande:1977rw,Barbieri:2006dq} is one of the simplest extensions of the SM, and corresponds to the limit of the 2HDM in which the previously-mentioned \zt~symmetry is \textit{exact} after EWSB. This ensures that there is no mixing between the SM-like doublet $\Phi_1$, and the \zt-odd one $\Phi_2$. Furthermore, it allows the model to accommodate a dark matter candidate -- namely the lightest \zt-odd scalar. Under the gauge and \zt~symmetries, the scalar potential is given by 
\begin{align}
\label{EQ:IDM_pot}
 \vtree_\text{IDM}=&\mu_1^2|\Phi_1|^2+\mu_2^2|\Phi_2|^2+\frac{\lambda_1}{2}|\Phi_1^2|^4+\frac{\lambda_2}{2}|\Phi_2^2|^4+\lambda_3|\Phi_1|^2|\Phi_2|^2+\lambda_4|\Phi_1^\dagger\Phi_2|^2+\frac{\lambda_5}{2}\left((\Phi_1^\dagger\Phi_2)^2+\hc\right)\,.
\end{align}
Following Ref.~\cite{Aoki:2013lhm}, we here decompose the two doublets in terms of mass eigenstates as 
\begin{equation}
 \Phi_1=\begin{pmatrix}
          G^+\\
          \frac{1}{\sqrt{2}}(v+h+i G)
        \end{pmatrix}\,,
\qquad
 \Phi_2=\begin{pmatrix}
          H^+\\
          \frac{1}{\sqrt{2}}(H+iA)
        \end{pmatrix}
       \,.
\end{equation}
where we use the same notations as in the 2HDM. Finally, from the above tree-level potential, we can derive field-dependent masses for the inert ($i.e.$ \zt-odd) scalars $H$, $A$, and $H^\pm$ as $m_\Phi^2(h)=\mu_2^2+\lambda_\Phi/2(v+h)^2$, where $\lambda_{H,A}=\lambda_3+\lambda_4\pm\lambda_5$ and $\lambda_{H^\pm}=\lambda_3$.

%%%%%%%%%%%%%%%%%%%%%%%%%%%%%%%%%%%%%%%%%%%%%%%%
%%%%%%%%%%%   EFFECTIVE POTENTIAL   %%%%%%%%%%%%
%%%%%%%%%%%%%%%%%%%%%%%%%%%%%%%%%%%%%%%%%%%%%%%%

\section{Effective-potential calculation of $\lambda_{hhh}$ at two loops}
\label{SEC:SETUP}

We investigate leading two-loop corrections to the effective Higgs trilinear coupling in the effective-potential approximation, which is equivalent to setting external momenta to zero in a diagrammatic calculation. We define our loop expansion of the effective potential $\veff$ as
\begin{equation}
\label{EQ:eff_pot_exp}
 \veff\equiv\vtree+\Delta\veff=\vtree+\kappa \vone + \kappa^2 \vtwo + \cdots\,,
\end{equation}
where $\kappa\equiv1/(16\pi^2)$ is the usual loop factor. 
While they miss potential threshold effects (shown at one loop for example in Ref. \cite{Kanemura:2004mg}), effective-potential computations are considerably simpler than diagrammatic ones and are sufficient for a first study of the magnitude of two-loop corrections. Furthermore, from past experience with scalar mass calculations we may expect the inclusion of momentum at two loops to give only subleading effects -- see $e.g.$ \cite{Martin:2004kr,Borowka:2014wla,Degrassi:2014pfa,Braathen:2017izn}. 

Normalising the effective Higgs trilinear coupling as $\lagr\supset-\frac16\lambda_{hhh}h^3$, the radiative corrections that it receives can be computed by taking derivatives of the effective potential as 
\begin{equation}
 \lambda_{hhh}=\frac{\partial^3\veff}{\partial h^3}\bigg|_\text{min}=\lambda_{hhh}^{(0)}+\Delta\lambda_{hhh}=\lambda_{hhh}^{(0)}+\kappa\delo\lambda_{hhh}+\kappa^2\delt\lambda_{hhh}+\cdots\,.
\end{equation}
In the scenarios without mixing in the scalar sector that we consider in section~\ref{SEC:Numerical}, the tree-level result $\lambda_{hhh}^{(0)}$ can be simply expressed in terms of $[M_h^2]_{V_\text{eff}}=\left[-\frac{1}{v}\frac{\partial}{\partial h}+\frac{\partial^2}{\partial h^2}\right]\veff\big|_\text{min.}$, the effective-potential (or curvature) mass of the lightest Higgs boson, as 
\begin{equation}
 \lambda_{hhh}=\frac{3[M_h^2]_{V_\text{eff}}}{v}+\mathcal{D}_3\Delta\veff\Big|_\text{min}\quad\text{where}\quad\mathcal{D}_3\equiv\frac{\partial^3}{\partial h^3}-\frac{3}{v}\left[-\frac{1}{v}\frac{\partial}{\partial h}+\frac{\partial^2}{\partial h^2}\right]\,.
\end{equation}
The above definition of the differential operator $\dlamb$ ensures that tadpole conditions are taken into account -- $i.e.$ the calculation is performed at the minimum of the loop-corrected potential. 

We follow the common choice of performing renormalisation \textit{before} taking derivatives of the potential, which allows us to make use of existing results for the effective potential \cite{Martin:2001vx}. The renormalised effective potential is calculated in terms of field-dependent (\msbar) tree-level masses, and therefore the results we find for $\lambda_{hhh}$ are also expressed in terms of \msbar-renormalised parameters. While theoretically consistent and simple, \msbar-scheme calculations may be plagued by large logarithmic contributions that appear because of the explicit renormalisation scale dependence, and furthermore it requires the inclusion of renormalisation group equations (RGEs) for all running parameters. Therefore, we choose to use an OS scheme instead and express our results in terms of physical parameters. For this purpose, we translate the relevant parameters, $i.e.$ all particle masses and the Higgs VEV, from their \msbar values $[M_h^2]_{V_\text{eff}},\,m_H,\,m_A,\,m_{H^\pm},\,m_t,\,v$ to OS ones $M_h,\,M_H,\,M_A,\,M_{H^\pm},\,M_t,\,\vphys=(\sqrt{2}G_F)^{-1/2}$ ($G_F$ being the Fermi constant), and we also include finite wave-function renormalisation (WFR) as
\begin{equation}
 \hat{\lambda}_{hhh}=\left(\frac{Z_h^\text{OS}}{Z_h^\msbar}\right)^{3/2}\lambda_{hhh}=\left(1+\frac32\delta Z_h^\text{OS}-\frac32\delta Z_h^\msbar\right)\lambda_{hhh}=\bigg(1+\frac32\frac{d}{dp^2}\Pi_{hh}(p^2)\big|_{p^2=M_h^2}\bigg)\lambda_{hhh}\,,
\end{equation}
where $\delta Z_h^\text{OS}$ and $\delta Z_h^\msbar$ are the WFR counterterms in the OS and \msbar schemes respectively, and $\Pi_{hh}(p^2)$ is the finite part of the Higgs self-energy evaluated at external momentum equal to $p^2$. We recall that the pole and curvature masses of the Higgs boson are related as (see $e.g.$ \cite{Degrassi:2012ry})
\begin{equation}
 M_h^2=[M_h^2]_{V_\text{eff}}+\Pi_{hh}(p^2=M_h^2)-\Pi_{hh}(p^2=0)\,.
\end{equation}

As our main concern is the size of the dominant two-loop BSM contributions to $\lambda_{hhh}$ due to the additional scalar states in the 2HDM and the IDM, we make the further approximation of neglecting contributions from the 125-GeV Higgs and the would-be Goldstone bosons, at both one- and two-loop orders, throughout the following. This will not impact our conclusions on the possible magnitude of two-loop effects, because these effects are maximal for large BSM-scalar masses, and this is also exactly when the validity of the approximation of neglecting the lighter-scalar masses is best. Moreover, as we work here in aligned scenarios of New Physics, corrections involving only masses of the Goldstone and 125-GeV-Higgs bosons are common with the SM, and will therefore drop out of the BSM deviations that we will present in the next section. 

Before turning to the two-loop computation and our new results, we briefly review here the effective-potential calculation of one-loop corrections to the Higgs trilinear coupling. The dominant terms in the one-loop effective potential, for both the 2HDM and the IDM, are \cite{Jackiw:1974cv} 
\begin{equation}
 \vone=-3m_t^4(h)\left(\llog m_t^2(h)-\frac32\right)+\sum_{\Phi=H,A,H^\pm}\frac{n_\Phi m_\Phi^4(h)}{4}\left(\llog m_\Phi^2(h)-\frac32\right)\,,
\end{equation}
where $m_t^2(h)$ and $m_\Phi^2(h)$ are the field-dependent masses of the top quark and of the extra scalars, respectively, and $n_\Phi$ is 1 for $H$ and $A$, and 2 for $H^\pm$ -- as mentioned earlier, we have neglected here the SM-like Higgs and Goldstone boson terms. 

One can then derive the leading one-loop contributions to $\lambda_{hhh}$ by using the operator $\dlamb$ \cite{Kanemura:2002vm,Senaha:2018xek}
\begin{equation}
\label{EQ:oneloop_res}
 \dlamb\vone\Big|_\text{min}=-\frac{48m_t^4}{v^3}+\sum_{\Phi=H,A,H^\pm}\frac{4n_\Phi m_\Phi^4}{v^3}\bigg(1-\frac{\mathcal{M}^2}{m_\Phi^2}\bigg)^3\,,
\end{equation}
where $\mathcal{M}$ is either $M$ in the 2HDM or $\mu_2$ in the IDM. 

\begin{figure}
\centering
 \includegraphics[width=.6\textwidth]{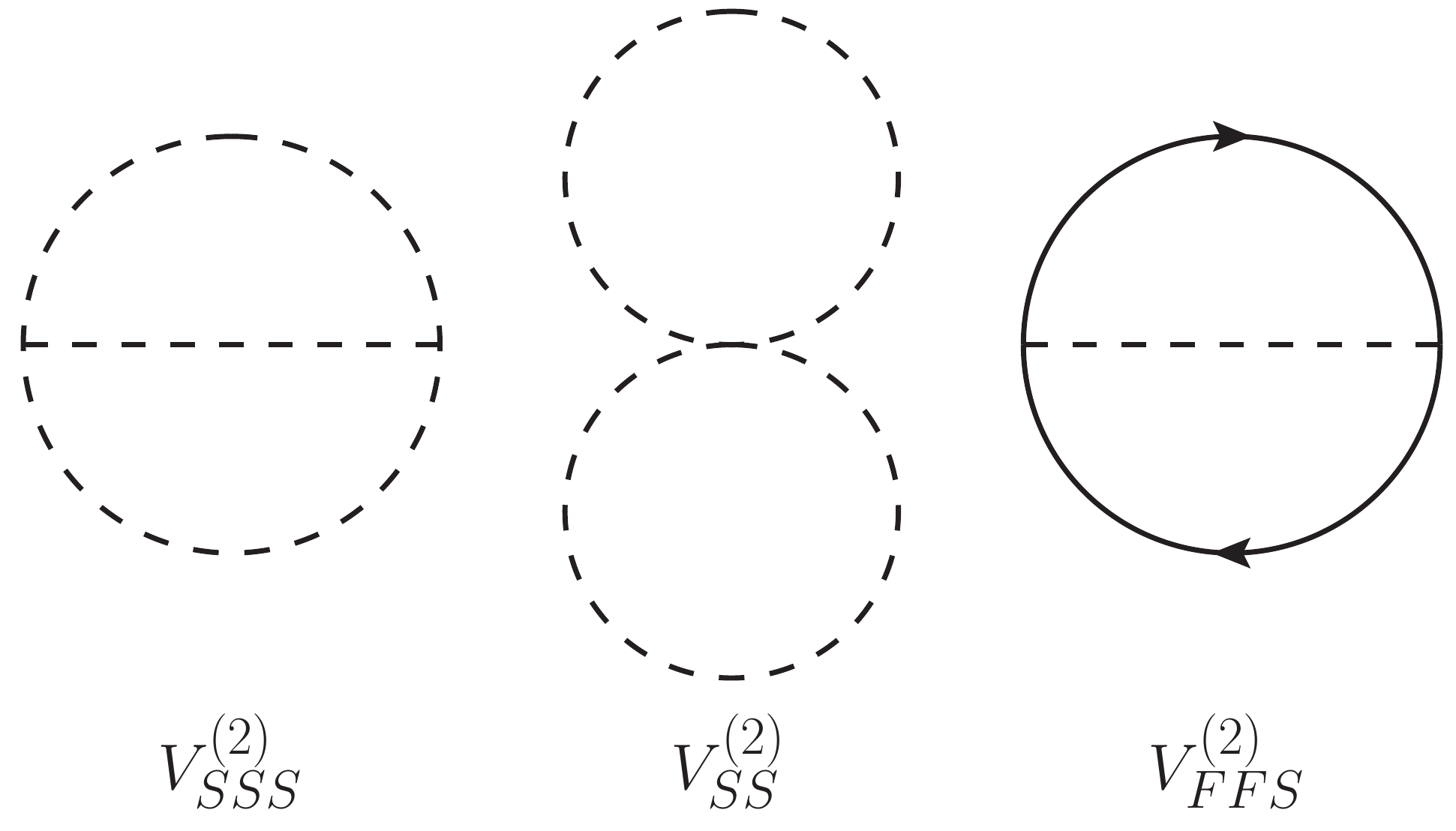}
 \caption{Topologies of diagrams with scalars and fermions contributing to the effective potential at two loops.}
 \label{FIG:diags}
\end{figure}

Corrections to the two-loop effective potential are obtained by calculating one-particle-irreducible vacuum bubble diagrams \cite{Jackiw:1974cv}. For our study, we expand the two-loop part of $\veff$ as $\vtwo=\vtwo_{SSS}+\vtwo_{SS}+\vtwo_{FFS}$, where each index $S$ or $F$ indicates a scalar or Dirac-fermion propagator -- the corresponding diagrams are shown in figure~\ref{FIG:diags}. Analytic expressions for each of these terms can be obtained in the Landau gauge and \msbar scheme for any renormalisable model using\footnote{Note that our notation differs slightly from that of Ref.~\cite{Martin:2001vx} because we work here with Dirac fermions, and not Weyl fermions, therefore our $\vtwo_{FFS}$ corresponds to the sum $\vtwo_{FFS}+\vtwo_{\bar{F}\bar{F}S}$ in \cite{Martin:2001vx}.} the results of Ref.~\cite{Martin:2001vx} (see also \cite{Martin:2018emo} for results with a general gauge fixing). These involve only two loop functions, namely the one-loop function $A$ and the two-loop sunrise integral $I$, for both of which complete expressions are given $e.g.$ in Refs. \cite{Ford:1992pn, Martin:2001vx, Martin:2003qz, Degrassi:2009yq}, and useful limits of $I$ with one or more mass arguments equal or vanishing can be found in Refs. \cite{Martin:2003qz,Braathen:2016cqe}.

At two loops, the \msbar to OS scheme conversion requires adding finite one-loop or two-loop shifts to the parameters that enter at one loop and tree level respectively. However for the latter, $i.e.$ the Higgs mass $[M_h^2]_{\veff}$ and VEV $v$, the two-loop shifts yield corrections to $\hat\lambda_{hhh}$ proportional to the 125-GeV Higgs mass and should hence be neglected in our approximation. Similarly, corrections involving two-loop WFR are also proportional to the 125-GeV Higgs mass and thus subleading. To summarise, we only need one-loop scheme translations for the Higgs VEV, the scalar masses, and the top quark mass, as well as one-loop finite WFR. 

Finally, before considering BSM corrections, we should mention also the case of the SM calculation, performed, for example, in Ref.  \cite{Senaha:2018xek}. Starting from the two-loop SM effective potential given in Ref. \cite{Ford:1992pn}, we obtain the same result as equation (11) of \cite{Senaha:2018xek} in terms of \msbar parameters. However, when translating that expression to the OS scheme, both using the results of Ref.~\cite{Degrassi:2012ry} as well as with a standalone calculation, we have
\begin{equation}
 \delt\hat\lambda_{hhh}=\frac{72M_t^4}{\vphys^3}\bigg(16g_3^2-\frac{13M_t^2}{\vphys^2}\bigg)\,.
\end{equation}
Our results do not agree with equation (12) of \cite{Senaha:2018xek} as we find for the numerical coefficient of the two-loop $M_t^6$ term $-936$ instead of $336$. We have furthermore checked that the numerical values for our OS expression and the \msbar one evaluated at renormalisation $Q=M_t$ are in excellent agreement.

%%%%%%%%%%%%%%%%%%%%%%%%%%%%%%%%%%%%%%%%%%%%%%%
%%%%%%%%%%%   Numerical examples   %%%%%%%%%%%%
%%%%%%%%%%%%%%%%%%%%%%%%%%%%%%%%%%%%%%%%%%%%%%%

\section{Numerical examples}
\label{SEC:Numerical}

\subsection{An aligned scenario with degenerate heavy scalars in the 2HDM}
\label{SEC:num_2HDM}

As our first numerical example, we consider a simplified scenario of the 2HDM where the additional scalars are degenerate in mass. This ensures that our calculations contain only three mass scales $\tilde M$, $M_\Phi$, and $M_t$ and allow relatively compact analytical expressions to be obtained. Furthermore, to avoid complications arising from mixing between $h$ and $H$, the CP-even mixing angle $\alpha$ is fixed\footnote{Note that, in principle, we should take into account radiative corrections to the alignment condition, but as was studied $e.g.$ in \cite{Braathen:2017izn} these are typically minute, and we will neglect these effects here. } as $\alpha=\beta-\pi/2$ to ensure alignment.

In the 2HDM, there are with respect to the SM fifteen new diagrams involving heavy scalars and top quarks that contribute to the effective potential at two loops, which we can write as $\vtwo_{hHH}$, $\vtwo_{hAA}$, $\vtwo_{hH^\pm H^\pm}$, $\vtwo_{HHH}$, $\vtwo_{HAA}$, $\vtwo_{HH^\pm H^\pm}$, $\vtwo_{HH}$, $\vtwo_{HA}$, $\vtwo_{HH^\pm}$, $\vtwo_{AA}$, $\vtwo_{AH^\pm}$, $\vtwo_{H^\pm H^\pm}$, $\vtwo_{ttH}$, $\vtwo_{ttA}$, and  $\vtwo_{tbH^\pm}$. Applying the operator $\dlamb$ to these effective-potential terms, we obtain
\begin{align}
 \delta^{(2)}\lambda_{hhh}=&\ \frac{16 m_\Phi^4}{v^5}\left(4+9\cot^22\beta\right) \left(1-\frac{M^2}{m_\Phi^2}\right)^4 \big[-2 M^2 - m_\Phi^2 + (M^2 + 2 m_\Phi^2) \llog m_\Phi^2\big]\nn\\
 &+ \frac{192 m_\Phi^6 \cot^22\beta}{v^5} \left(1-\frac{M^2}{m_\Phi^2}\right)^4 \big[1+2\llog m_\Phi^2\big]\nn\\
 &+\frac{96  m_\Phi^4m_t^2 \cot^2\beta}{ v^5}\left(1-\frac{M^2}{m_\Phi^2}\right)^3 \big[-1 + 2 \llog m_\Phi^2 \big]+\mathcal{O}\left(\frac{m_\Phi^2m_t^4}{v^5}\right)\,.
\end{align}
in terms of the \msbar-renormalised parameters -- $m_\Phi$ being an \msbar-scheme degenerate mass for the heavy scalars $H,\,A,\,H^\pm$ -- and with $\llog x\equiv\log (x/Q^2)$, $Q$ being the renormalisation scale. The complete expression of the third derivative of the heavy scalar and top quark sunrise $\vtwo_{FFS}$ diagrams is rather long, so for brevity we only write here the leading $\mathcal{O}(m_\Phi^4m_t^2/v^5)$ term (while we use the complete result for our following numerical investigation). We performed some consistency checks of these results by: $(i)$ verifying that the $\log Q^2$ dependence of the total result for $\lambda_{hhh}$ is eliminated when including the running of all parameters appearing at lower orders; $(ii)$ confirming that each of the terms in this \msbar expression independently decouples when taking the limit $M\to\infty$. The latter can be understood as each term is proportional to
\begin{equation}
\label{EQ:decoup}
 (m_\Phi^2)^{n-1}\left(1-\frac{M^2}{m_\Phi^2}\right)^n\underset{m_\Phi^2=M^2+\tilde\lambda v^2}{=}\frac{(\tilde\lambda v^2)^n}{M^2+\tilde\lambda v^2}\,,
\end{equation}
with $n=3$ or $4$, and where $\tilde\lambda$ denotes some combination of Lagrangian scalar quartic couplings. We should expect to observe decoupling of the BSM corrections when taking $M\to\infty$ while keeping $\tilde\lambda v^2$ finite, so that the additional scalar masses go to infinity without the calculation entering the non-decoupling regime associated with large scalar quartic couplings, which would cause perturbativity to be lost. Here, when taking the limit $M\to\infty$ with $\tilde\lambda v^2$ fixed our expressions do indeed decouple properly, as can be seen from eq.~(\ref{EQ:decoup}). 

Instead of \msbar-renormalised parameters, we prefer to work in terms of physical parameters, and therefore we now convert our expressions to the OS scheme. For the masses of the top quark and the heavy scalars, this simply requires shifting the masses in the one-loop corrections given in eq.~(\ref{EQ:oneloop_res}) by the corresponding self-energies. At this point, we should emphasise that the scenarios where the heavy scalars have degenerate \msbar masses or OS masses correspond to different points in the parameter space of the 2HDM, because the radiative corrections that relate \msbar and OS masses are not the same for the different scalars. Keeping this in mind, we choose however to consider parameter points for which the scalars $H$, $A$, and $H^\pm$ have a common physical mass $M_\Phi$, after the conversion of our results to the OS scheme. Then, for $\tan\beta$, we do not need to perform any conversion, as this parameter only enters the calculation at two loops. 

Finally, the treatment of $M^2$ is more subtle, as we will discuss now. When working at one-loop order, one may find decoupling of the effects of the heavy-scalar loops in the OS scheme result in the limit $M\to\infty$ when using a relation $M_\Phi^2=M^2+\tilde\lambda v^2$, with $M_\Phi$ renormalised in the OS scheme and $M$ in the \msbar scheme \cite{Kanemura:2004mg}. However, when going to two-loop order this is not the case any more, and one needs to relate parameters that appear at one-loop order in different schemes with a one-loop equation, so as not to miss two-loop order effects.  We have checked that decoupling does occur if we consistently use a one-loop relation between $M_\Phi$ and $M$ in our results -- this is essentially equivalent to using expressions with the heavy scalar masses renormalised in the \msbar scheme. Nevertheless, this situation motivates devising an ``on-shell'' renormalisation\footnote{We write here ``on-shell'' with inverted commas, as we are not actually relating $\tilde M$ to some physical observable.} condition for the soft-breaking mass, which we then denote $\tilde M$, that would make decoupling apparent when using a relation of the form $M_\Phi^2=\tilde M^2+\tilde\lambda \vphys^2$. Furthermore, we still have the freedom to choose this renormalisation condition in such a way that it ensures the complete cancellation of all $\llog m_\Phi^2$ terms in $\delta^{(2)}\lambda_{hhh}$. We emphasise here that $\tilde M$ is the OS-renormalised value of the soft-breaking scale of the $\mathbb{Z}_2$ symmetry of the 2HDM, and that therefore it is the parameter that governs the possibility of decoupling of the additional states in the extended Higgs sector. 

With this choice, we can derive the finite ``OS'' counterterm $\delta^\text{OS}M^2$ for $M^2$ -- defined so that $\tilde M^2=M^2+\delta^\text{OS}M^2$ -- and we obtain at the one-loop order
\begin{align}
\label{EQ:deltaMOS}
 \delta^\text{OS}M^2=-\kappa\bigg(&\frac{12(M^2-m_\Phi^2)M^2\cot^22\beta}{v^2}[\llog m_\Phi^2-1]\nn\\
                     &+\frac{3M^2m_t^2\cot^2\beta}{v^2}\Big[B_0(m_\Phi^2,m_t^2,m_t^2)+B_0(m_\Phi^2,0,m_t^2)\Big]\bigg)\,,
\end{align}
where $B_0$ is the (finite part of the) usual Passarino-Veltman function \cite{Passarino:1978jh}. Our final OS scheme result for $\delt\hat\lambda_{hhh}$ is then
\begin{align}
\label{EQ:2hdm_OS_res}
 \delt\hat\lambda_{hhh}=&\ \frac{48 M_\Phi^6}{\vphys^5}\left(1-\frac{\tilde M^2}{M_\Phi^2}\right)^4 \left\{4+3\cot^22\beta\left[3-\frac{\pi}{\sqrt{3}}\left(\frac{\tilde M^2}{M_\Phi^2} + 2\right)\right]\right\}\nn\\
                          &+\frac{576 M_\Phi^6 \cot^22\beta}{\vphys^5} \left(1-\frac{\tilde M^2}{M_\Phi^2}\right)^4+\frac{288 M_\Phi^4M_t^2 \cot^2\beta}{\vphys^5}\left(1-\frac{\tilde M^2}{M_\Phi^2}\right)^3\nn\\
                          &+\frac{168M_\Phi^4M_t^2}{\vphys^5}\left(1-\frac{\tilde M^2}{M_\Phi^2}\right)^3-\frac{48M_\Phi^6}{\vphys^5}\left(1-\frac{\tilde M^2}{M_\Phi^2}\right)^5+\mathcal{O}\left(\frac{M_\Phi^2M_t^4}{\vphys^5}\right)\,,
\end{align}
where the terms on the third line come from finite WF and VEV renormalisation. 

\begin{figure}
\centering
 \includegraphics[width=.495\textwidth]{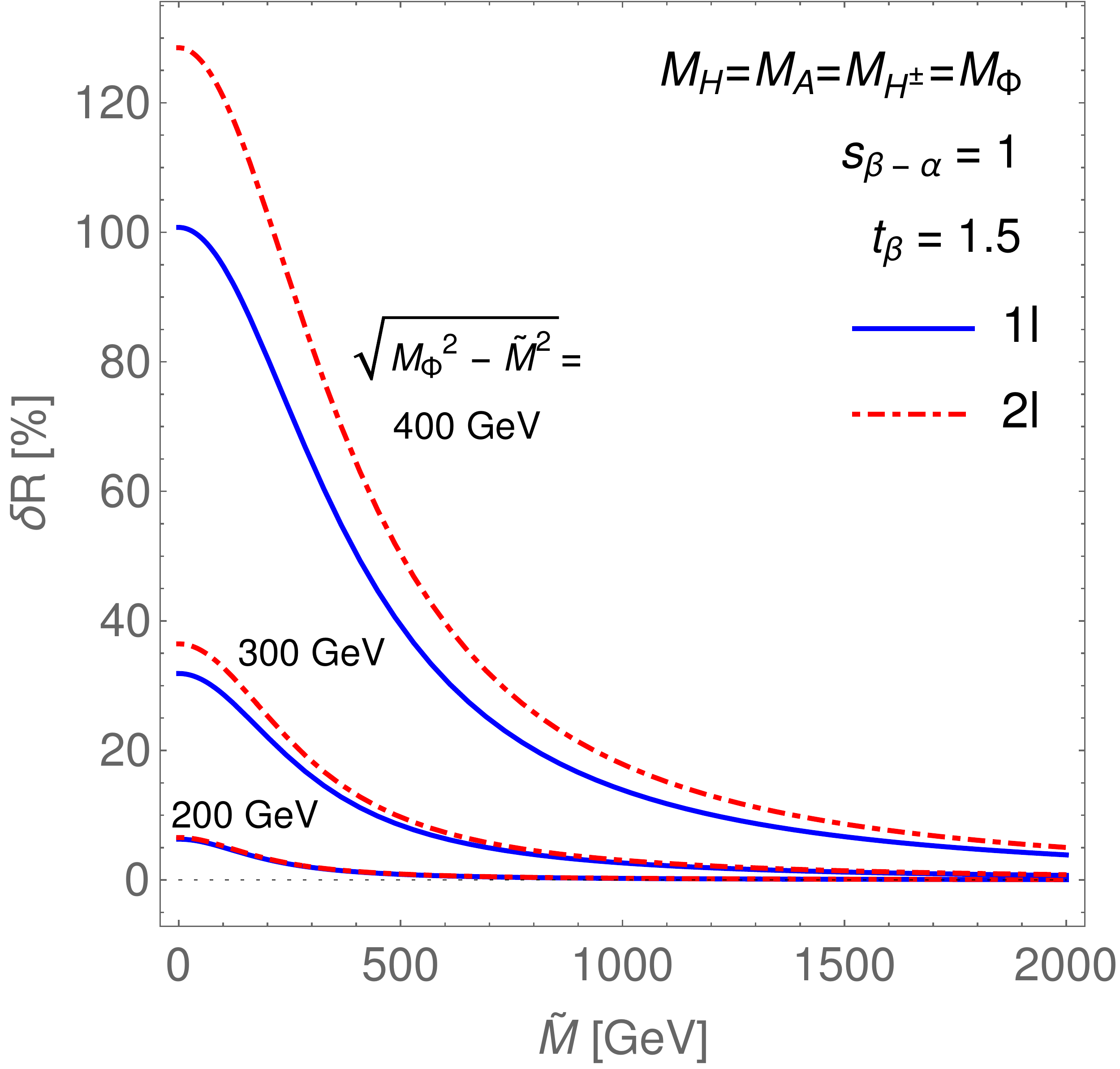}
 \includegraphics[width=.49\textwidth]{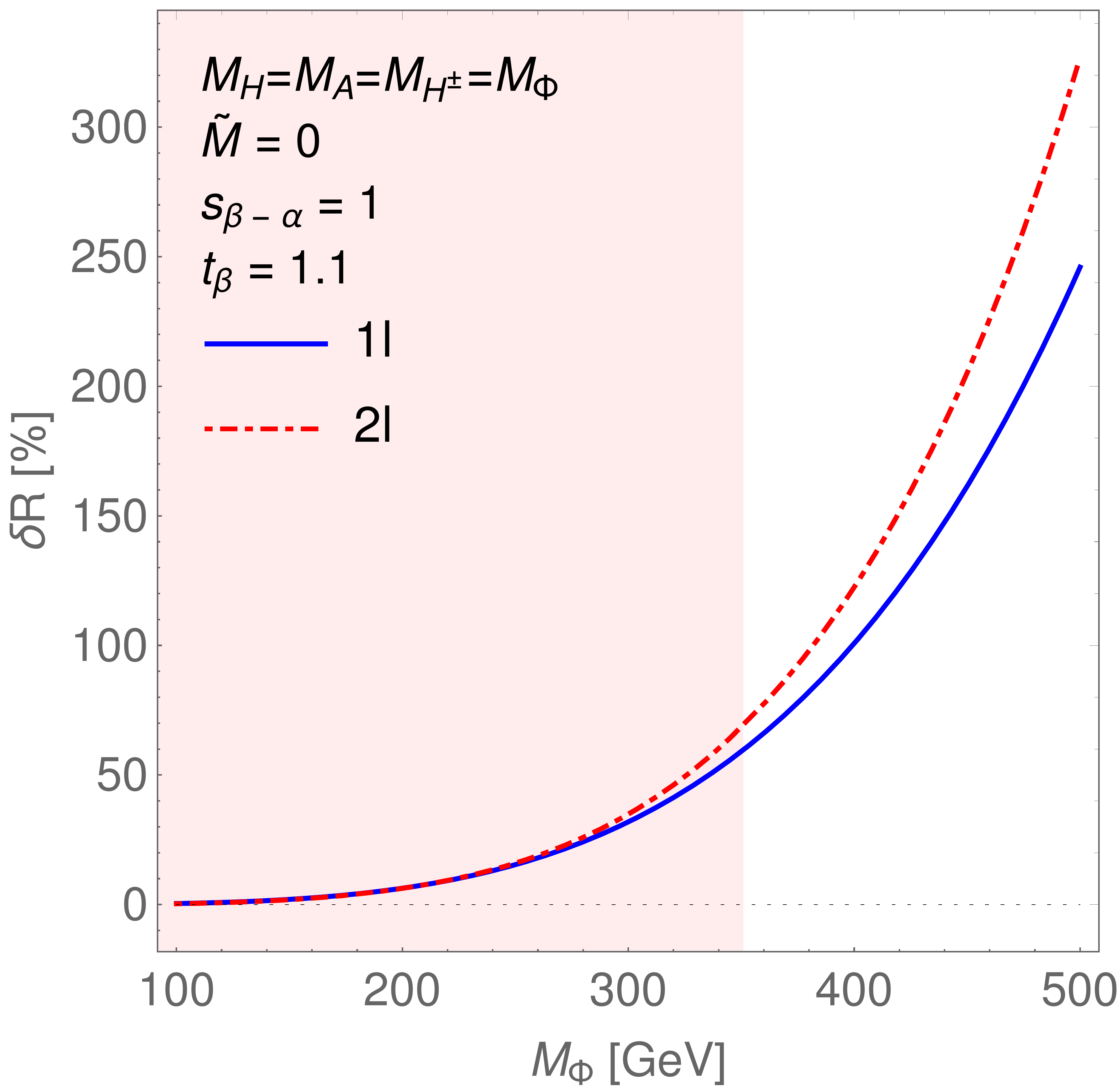}
 \caption{Illustrations of our results for the deviation $\delta R$ of $\hat\lambda_{hhh}$ computed in the 2HDM with respect to its SM prediction, with $\delta R\equiv\Delta\hat\lambda_{hhh}^\text{2HDM}/\hat\lambda_{hhh}^\text{SM}=\hat\lambda_{hhh}^\text{2HDM}/\hat\lambda_{hhh}^\text{SM}-1$. 
          \textbf{(Left side)}: Decoupling behaviour of the BSM contributions to $\hat\lambda_{hhh}$, shown by plotting $\delta R$ at one loop (\textit{solid blue curve}) and two loops (\textit{red dot-dashed curve}) as a function of $\tilde M$. The degenerate pole mass of the additional scalars $M_\Phi$ is taken to be $M_\Phi^2=\tilde M^2+\tilde\lambda\vphys^2$, with $\tilde\lambda\vphys^2=(200\text{ GeV})^2,\,(300\text{ GeV})^2,\,(400\text{ GeV})^2$, and we fix $\tan\beta=1.5$. 
          \textbf{(Right side)}: $\delta R$ computed at one loop (\textit{solid blue curve}) and two loops (\textit{red dot-dashed curve}) as a function of $M_\Phi$, for the maximal non-decoupling case of $\tilde M=0$, and with $\tan\beta=1.1$. The light-red shaded region shows the values of the additional scalar masses currently excluded by experimental searches. It should be noted that this constraint on $M_\Phi$ can be weakened for $\tilde{M}>0$.}
 \label{FIG:2HDM_res}
\end{figure}

These two-loop corrections indeed decouple explicitly when taking the limit $\tilde M\to\infty$ with $\tilde\lambda\vphys^2$ fixed. An example of this is shown in the left side of figure~\ref{FIG:2HDM_res}, where we plot the deviation $\delta R\equiv\hat\lambda_{hhh}^\text{2HDM}/\hat\lambda_{hhh}^\text{SM}-1$ of $\lambda_{hhh}$ calculated in the 2HDM with respect to the SM prediction as a function of the OS-renormalised $\tilde M$, at one- and two-loop orders, for different fixed values of $\sqrt{M_\Phi^2-\tilde M^2}$ and for $\tan\beta=1.5$ -- other physical inputs being taken from the PDG \cite{Tanabashi:2018oca}. 

A comment about experimental constraints on the 2HDM parameter space should also be made at this point. Indeed, the allowed values of the charged Higgs mass and of $\tan\beta$ can be constrained by searches of charged and neutral scalars at the LHC as well as by results on flavour observables -- some detailled discussions can be found, for example, in Refs.~\cite{Misiak:2017bgg, Arbey:2017gmh}. Furthermore, with the \texttt{Mathematica} package \texttt{SARAH} \cite{Staub:2008uz,Staub:2009bi,Staub:2010jh,Staub:2012pb,Staub:2013tta,Staub:2015kfa}, we have created a \texttt{SPheno} \cite{Porod:2003um,Porod:2011nf} spectrum generator for a type-I 2HDM\footnote{Other types of 2HDMs, in particular type II and type Y, are more severely constrained by flavour observables -- see $e.g.$ Ref.~\cite{Misiak:2017bgg}. }, which allowed us to check with \texttt{HiggsBounds-5.3.2beta} \cite{Bechtle:2008jh,Bechtle:2011sb,Bechtle:2013wla,Bechtle:2015pma} that for $\tilde M=0$ and $\tan\beta=1.1$ ($\tan\beta=1.5$), BSM scalar masses above $350\text{ GeV}$ ($355\text{ GeV}$) are not currently excluded by experimental searches. 

The other interesting limit to consider with our analytical results is the case of maximal non-decoupling effects that we obtain for $\tilde M=0$. We illustrate the non-decoupling behaviour of the BSM corrections to $\hat\lambda_{hhh}$ in the right side of figure~\ref{FIG:2HDM_res}, where we show the same deviation $\delta R$ as in the left-side plot but now as a function of the heavy scalar (degenerate) pole mass $M_\Phi$, in the case of $\tilde M=0$ (to enhance as much as possible the non-decoupling effects) and $\tan\beta=1.1$. From essentially negligible before the one-loop corrections for $M_\Phi\lesssim\vphys$, the two-loop contributions become as large as 80\% for $M_\Phi=500\text{ GeV}$ -- the one-loop deviation is then 250\%. One should note that the value $M_\Phi=500\text{ GeV}$ is close to the upper limit on $M_\Phi$ allowed by the criterion of tree-level unitarity -- using \cite{Kanemura:1993hm}, we find this limit to be $M_\Phi\simeq600\text{ GeV}$ for $\tan\beta=1.1$ and $\tilde M=0$. For $M_\Phi=400\text{ GeV}$, well below the bound from perturbative unitarity, the BSM contributions cause a deviation of $\hat\lambda_{hhh}^\text{2HDM}$ with respect to its SM prediction of 101\% at one loop, and of a further 22\% at two loops. Finally, we should mention also the dependence on $\tan\beta$ that appears at two loops in $\hat\lambda_{hhh}$, even in the alignment limit that we have considered. In particular, the effect of $\tan\beta$ is largest in the scalar contributions to $\delt\hat\lambda_{hhh}$ (the first two terms in eq.~(\ref{EQ:2hdm_OS_res})), because of their $\cot^22\beta$ dependence, and hence these terms are greatly enhanced when $\tan\beta$ increases. However, we observe that the perturbative expansion is not broken -- $i.e.$ two-loop corrections to $\hat\lambda_{hhh}$ remain smaller than their one-loop counterparts -- at least while perturbative unitarity conditions are not violated. To illustrate this, we consider two example points. For the first one, we take $M_\Phi=400\text{ GeV}$ and $\tilde M=0$, and the criterion of tree-level unitarity then implies \cite{Kanemura:1993hm} an upper bound $\tan\beta\lesssim 1.7$. With this maximal value of $\tan\beta$, the one- and two-loop deviations of $\hat\lambda_{hhh}$ from BSM contributions are respectively 101\% and 34\%. We fix for the second example point $M_\Phi=250\text{ GeV}$ and $\tilde M=0$, which gives the bound $\tan\beta\lesssim 2.8$, and in turn we obtain for the deviation of $\hat\lambda_{hhh}^\text{2HDM}$ from $\hat\lambda_{hhh}^\text{SM}$ at one- and two-loop orders respectively 15\% and 6\%. It therefore appears that, under the criterion of tree-level unitarity, the two-loop BSM corrections to the Higgs trilinear coupling can become at most $\mathcal{O}(30-40\%)$ of the one-loop ones.

\subsection{A dark-matter-inspired scenario in the IDM}
\label{SEC:num_IDM}

The second case that we consider is a scenario of the IDM, already studied in Ref.~\cite{Senaha:2018xek}, in which the additional inert scalar $H$ is light ($M_H\simeq M_h/2\ll M_{A,H^\pm}$) and becomes a DM candidate. The leading two-loop corrections to $\lambda_{hhh}$ are then due to the pseudoscalar and charged Higgs bosons, and in order to maximise the size of the radiative corrections, we set the mass parameter $\mu_2$ to zero throughout this section. 

In this scenario, there are eight BSM diagrams that give contributions to $\vtwo$, namely $\vtwo_{hAA}$, $\vtwo_{hH^\pm H^\pm}$, $\vtwo_{AH^\pm G^\pm}$, $\vtwo_{HAG}$, $\vtwo_{HH^\pm G^\pm}$, $\vtwo_{AA}$, $\vtwo_{AH^\pm}$, and $\vtwo_{H^\pm H^\pm}$. Only the first two of these were included in Ref.~\cite{Senaha:2018xek}, and we find agreement between our results for these and equation (16) of~\cite{Senaha:2018xek}. Taking into account the other of the above diagrams, we present here for the first time the complete $\mathcal{O}(M_\Phi^6/\vphys^5)$ and $\mathcal{O}(\lambda_2M_\Phi^4/\vphys^3)$ contributions to $\delta^{(2)}\hat\lambda_{hhh}$ (here by $M_\Phi$ we mean $M_A$, $M_{H^\pm}$, or some combination of the two). After conversion to the OS scheme, and inclusion of finite WF and VEV renormalisation, these read
\begin{align}
\label{EQ:IDM_OSres}
 \delta^{(2)}\hat\lambda_{hhh}=&\frac{6 \lambda_2}{\vphys^3}\big(3M_A^4+4M_A^2 M_{H^\pm}^2+8M_{H^\pm}^4\big)+\frac{60 (M_A^6+2M_{H^\pm}^6)}{\vphys^5}+\frac{24(M_A^2-M_{H^\pm}^2)^2(M_A^2+M_{H^\pm}^2)}{\vphys^5}\nn\\
                                     &\hspace{-.2cm}+\frac{24M_t^4(M_A^2+2M_{H^\pm}^2)}{\vphys^5}+\frac{42M_t^2(M_A^4+2M_{H^\pm}^4)}{\vphys^5}-\frac{2(M_A^4+2M_{H^\pm}^4)(M_A^2+2M_{H^\pm}^2)}{\vphys^5}\,.
\end{align}
We emphasise that, as $\lambda_2$ only appears in the calculation of $\lambda_{hhh}$ at two-loop order, we do not need here to specify a choice of renormalisation scheme for it, as opposed to the inert scalar masses, which appear at one-loop. Moreover, we expect from our result for $\tilde M$ in the previous section that the tree-level (\msbar) condition $\mu_2=0$ will also hold in the OS scheme (see eq.~(\ref{EQ:deltaMOS})). Interestingly, one may notice that although the inert scalars do not couple directly to the top quark, terms involving both the top and inert-scalar masses do appear at two loops through the interplay of one-loop scalar contributions to the Higgs WFR with the one-loop top quark correction to $\lambda_{hhh}$ (and vice versa). 

At this point, we can also quantify the effects of the new diagrams that we include here for the first time. When expressed in the OS scheme, the contributions arising from the two sunrise diagrams $\vtwo_{hAA}$ and $\vtwo_{hH^\pm H^\pm}$ read, respectively, $48M_A^6/\vphys^5$ and $96M_{H^\pm}^6/\vphys^5$. We then find that the corrections to $\delta^{(2)}\lambda_{hhh}$ obtained from $\vtwo_{HAG}$ and $\vtwo_{HH^\pm G^\pm}$ amount to 25\% of the aforementioned results, while the terms coming from the last sunrise diagram $\vtwo_{AH^\pm G^\pm}$ only give significant effects when $M_A$ and $M_{H^\pm}$ are not degenerate -- $e.g.$ for $M_A=2M_{H^\pm}$ these terms result in an additional 25\% positive correction to $\delta^{(2)}\lambda_{hhh}$ compared to the result of Ref.~\cite{Senaha:2018xek}. Moreover, the corrections from the other diagrams $\vtwo_{AA}$, $\vtwo_{AH^\pm}$, and $\vtwo_{H^\pm H^\pm}$ involving $\lambda_2$ can also become large, as we will see now.

\begin{figure}
\centering
 \includegraphics[width=.75\textwidth]{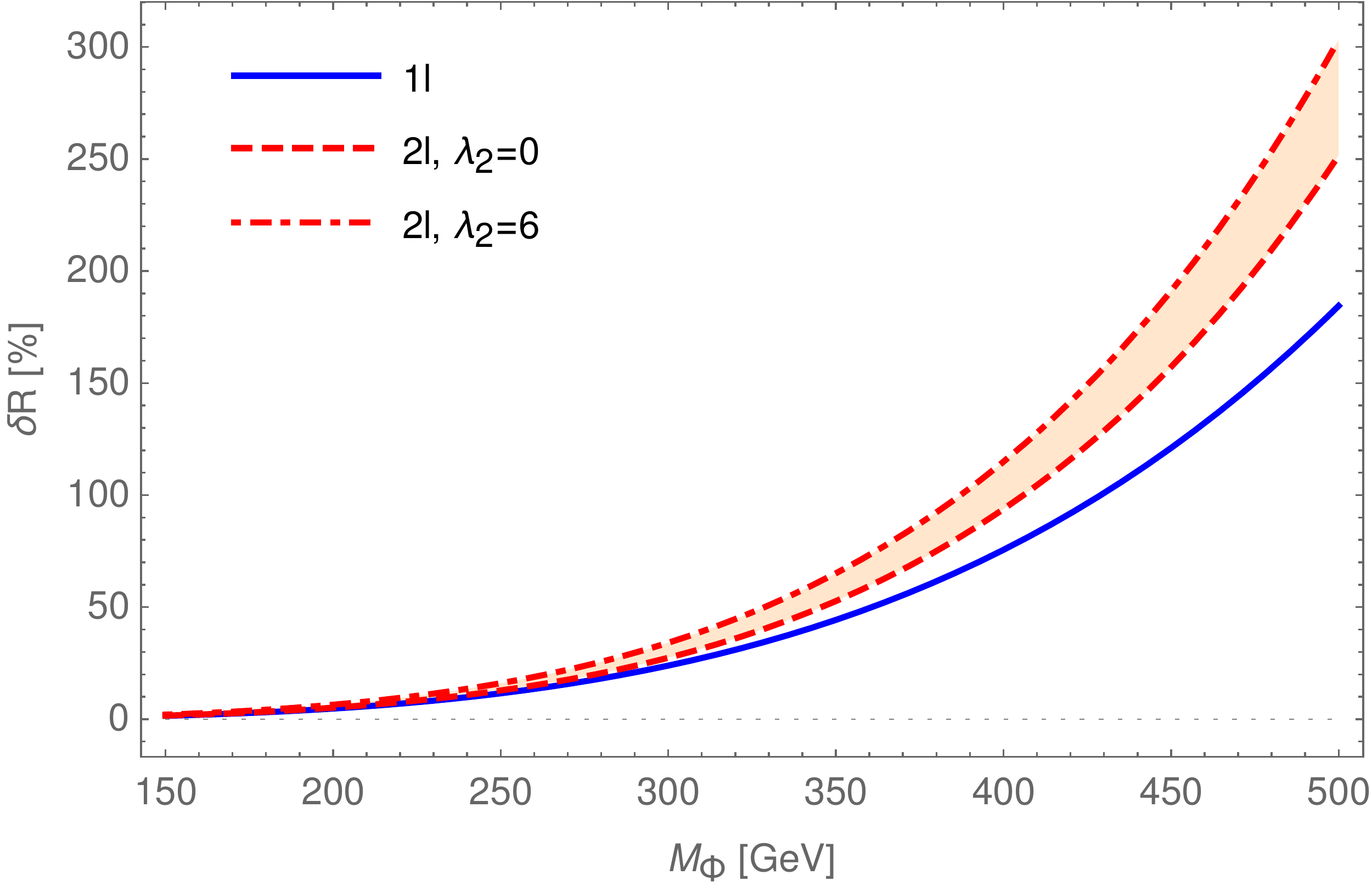}
 \caption{Deviation $\delta R$ of the Higgs trilinear coupling calculated in the IDM ($\hat\lambda_{hhh}^\text{IDM}$) with respect to the SM ($\hat\lambda_{hhh}^\text{SM}$) -- $i.e.$ $\delta R=\hat\lambda_{hhh}^\text{IDM}/\hat\lambda_{hhh}^\text{SM}-1$ -- as a function of the degenerate pole masses of the pseudoscalar and charged Higgses $M_\Phi=M_A=M_{H^\pm}$, at one loop (solid blue curve) and two loops (dashed and dot-dashed red curves). For the two-loop results, the dashed and dot-dashed correspond to different values of the inert doublet quartic coupling $\lambda_2$, respectively, $\lambda_2=0$ and $6$. We recall that we take $\mu_2=0$ (to maximise the non-decoupling effects), and that we neglect contributions from $h$ and $H$, as we assume $M_h,M_H\ll M_\Phi$.}
 \label{FIG:IDM_plot}
\end{figure}

We illustrate our numerical results in figure~\ref{FIG:IDM_plot}, where we plot the deviation $\delta R$ of the Higgs trilinear coupling $\hat\lambda_{hhh}^\text{IDM}$ with respect to its SM prediction $\hat\lambda_{hhh}^\text{SM}$, as a function of the pole masses of the heavy scalars, which we take to be equal -- $i.e.$ $M_A=M_{H^\pm}=M_\Phi$ -- both for convenience and to keep the $\rho$ parameter close to 1. One should note that this implies that the third term in the above equation~(\ref{EQ:IDM_OSres}), which corresponds to the $\vtwo_{AH^\pm G^\pm}$ diagram, vanishes. The impact of the non-vanishing sunrise diagrams -- $\vtwo_{hAA}$, $\vtwo_{hH^\pm H^\pm}$, $\vtwo_{HAG}$, and $\vtwo_{HH^\pm G^\pm}$ -- is given by the difference between the solid blue (one-loop) and dashed red (two-loop, with $\lambda_2=0$) curves in figure~\ref{FIG:IDM_plot}. We also try to evaluate the maximal possible size of the contributions proportional to $\lambda_2$ -- $i.e.$ coming from $\vtwo_{AA}$, $\vtwo_{AH^\pm}$, $\vtwo_{H^\pm H^\pm}$ -- under the constraint of perturbative unitarity, and for this purpose evaluate $\delta R$ at two loops for $\lambda_2=6$; note that we do not take $\lambda_2$ larger because we find the tree-level unitarity conditions to be violated \cite{Kanemura:1993hm,Akeroyd:2000wc} for $M_\Phi=500\text{ GeV}$, $\mu_2=0$ and $\lambda_2\simeq 6.5$. We note also that, in this type of IDM scenario with $H$ as a DM candidate of mass $M_H\simeq M_h/2$, the mass range that we consider here for $M_\Phi=M_A=M_{H^\pm}$ is not constrained by collider and DM searches \cite{Dercks:2018wch}. 

For $M_\Phi=400\text{ GeV}$, $\hat\lambda_{hhh}^\text{IDM}$ deviates at one loop by about $76\%$ with respect to its SM prediction, while the sunrise diagrams give a further enhancement of $\sim18\%$ and the remaining two-loop diagrams (with $\lambda_2$) another $\sim21\%$. On the one hand one may notice that, in relative size, the corrections from the sunrise diagrams grow significantly faster than the one-loop and two-loop $\lambda_2$-dependent ones, which can be understood from their expression proportional to $M_\Phi^6$ as opposed to $M_\Phi^4$ for the others. On the other hand, the remaining two-loop diagrams can potentially give large contributions for low $M_\Phi$, if $\lambda_2$ is large, because of their large combinatorial factor, but their relative importance diminishes for increasing $M_\Phi$. To summarise, similarly to what we found in the 2HDM, the two-loop corrections in the IDM remain smaller than the one-loop ones, meaning that the perturbative expansion is not breaking down, at least as long as perturbative unitarity is fulfilled.

Before concluding, we point out that while the coupling $\lambda_2$ is quite difficult to access experimentally as it is the quartic coupling between inert scalars (see the potential in eq.~(\ref{EQ:IDM_pot})), a precise measurement of $\lambda_{hhh}$ could allow us to obtain some information about the value of $\lambda_2$. We can expect this observation to hold also for couplings of the Higgs boson with other particles, such as for example the $h\gamma\gamma$ or $hZZ$ couplings, because the quartic coupling $\lambda_2$ should appear in internal scalar loops therein as well.

%%%%%%%%%%%%%%%%%%%%%%%%%%%%%%%%%%%%
%%%%%%%%%%%   Summary   %%%%%%%%%%%%
%%%%%%%%%%%%%%%%%%%%%%%%%%%%%%%%%%%%

\section{Summary}
\label{SEC:Conclusion}

We have discussed the magnitude of the deviation of the Higgs trilinear coupling $\lambda_{hhh}$ from its SM prediction at two loops in two models with extended Higgs sectors, namely the 2HDM  -- for which we have obtained for the first time leading two-loop corrections to $\lambda_{hhh}$ -- and the IDM -- where we have improved on the existing results of Ref.~\cite{Senaha:2018xek}. We have performed calculations both in the \msbar and the on-shell schemes using the effective-potential method. In the cases where it was possible, we have compared our expressions with existing works in the literature, explaining the origin of some differences in the SM and IDM. We also devised a new ``on-shell'' renormalisation prescription for the soft-breaking scale $\tilde M$ to maintain explicitly the decoupling of the two-loop corrections for $M_\Phi^2=\tilde M^2+\tilde\lambda\vphys^2$ with $\tilde M\to\infty$ and fixed $\tilde\lambda\vphys^2$. 

In the two models we studied, we found new dependences of $\lambda_{hhh}$ on parameters -- respectively $\tan\beta$ in the aligned 2HDM and $\lambda_2$ in the IDM -- entering the calculation only from two loops, and which may cause large enhancements of $\delt\hat\lambda_{hhh}$. 
However, we have shown that, provided one considers parameter points within the region of parameter space allowed under the criterion of tree-level unitarity, the two-loop corrections to $\lambda_{hhh}$ do not grow out of control and remain smaller than the one-loop effects. When expressed in terms of OS-scheme parameters, our new two-loop contributions (moderately) enhance the non-decoupling effects appearing at one-loop, but we should emphasise that we do not obtain new large -- $i.e.$ $\mathcal{O}(100\%)$ or so -- corrections at two loops. The typical size that we find for the two-loop corrections in the OS scheme, up to $\mathcal{O}(\sim 20\%)$ of the one-loop corrections, implies, on the one hand, that higher-order contributions do not change the existence of the non-decoupling effects observed from one loop. But on the other hand, it also means that in the perspective of precise measurements of the Higgs trilinear coupling, a careful theoretical calculation of $\lambda_{hhh}$ -- including radiative corrections beyond one loop -- will be necessary. 

Details about the calculations and discussions in this letter, and in particular complete expressions for a 2HDM, will be shown elsewhere \cite{BK:2019xxx}.

%%%%%%%%%%%%%%%%%%%%%%%%%%
%%%  Acknowledgments  %%%%
%%%%%%%%%%%%%%%%%%%%%%%%%%

\begin{acknowledgments}
 We are indebted to Kodai Sakurai for numerous useful discussions. 
 We would also like to thank Benjamin Fuks, Mark Goodsell, Kentarou Mawatari, and Sebastian Pa\ss{}ehr for helpful discussions. 
 This work is, in part, supported by Grant-in-Aid for Scientific Research on Innovative Areas, the Ministry of Education, Culture, Sports, Science and Technology, No. 16H06492 and No. 18H04587, and Grant H2020-MSCA-RISE-2014, No. 645722 (Non Minimal Higgs). 
 This work is also supported in part by JSPS KAKENHI Grant No. A18F180220. 
\end{acknowledgments}

\appendix

\bibliographystyle{utphys}
\bibliography{trilinear}

\end{document}